\documentclass[12pt]{article}
\usepackage[english,german]{babel}

\usepackage[inner=3.00cm,outer=3.00cm,top=2.5cm,bottom=2.5cm,includeheadfoot]{geometry}

\usepackage[percent]{overpic}
\usepackage{lscape}
\usepackage{amsmath}
\usepackage{bbm}
\usepackage{amssymb}
\usepackage{xcolor}
\usepackage{appendix}
\usepackage{rotating}
\usepackage[percent]{overpic}
\usepackage{longtable}
\usepackage{sidecap}

\begin{document}

\selectlanguage{german}


%
%

\title{Complete experiments in pseudoscalar meson photoproduction
}

\author{Yannick Wunderlich \\ Helmholtz-Institut f"ur Strahlen- und Kernphysik,\\ Universit"at Bonn \\ Nussallee 14-16,
 53115 Bonn, Germany 
}



\date{October 14, 2015}


\maketitle


\selectlanguage{english}

\begin{abstract}
The problem of extracting photoproduction amplitudes uniquely from so called complete experiments is discussed. This problem can be considered either for the extraction of full production amplitudes, or for the determination of multipoles. Both cases are treated briefly. Preliminary results for the fitting of multipoles, as well as the determination of their error, from recent polarization measurements in the $\Delta$-region are described in more detail.
\end{abstract}


\pagenumbering{arabic}

\section{Introduction to the formalism} \label{sec:Introduction}

For the photoproduction of a single pseudoscalar mesons, i.e. $\gamma N \longrightarrow \mathcal{P} B$, it can be shown that the most general expression for the reaction amplitude, with spin and momentum variables specified in the center of mass frame (CMS), reads (cf. the work by CGLN \cite{CGLN})
\begin{equation}
  F_{\mathrm{CGLN}} = i  \vec{\sigma} \cdot \hat{\epsilon}\; F_{1} + \vec{\sigma} \cdot \hat{q}
  \; \vec{\sigma} \cdot \hat{k} \times \hat{\epsilon}\; F_{2} + i  \vec{\sigma} \cdot \hat{k}\; \hat{q}
  \cdot \hat{\epsilon}\; F_{3} + i \vec{\sigma} \cdot \hat{q} \; \hat{q} \cdot \hat{\epsilon}\; F_{4}\,.
  \label{eq:DefCGLN}
 \end{equation}
Each spin-momentm structure in this expansion is multiplied by a complex function depending on the total energy $W$ and meson scattering angle $\theta$ 
\begin{table}[ht]
\centering
\caption{The 16 polarization observables accessible in pseudoscalar meson photoproduction (for a more elaborate version of this Table, cf. [2]).}
{\begin{tabular}{c|c|ccc|ccc|cccc}
\multicolumn{12}{c}{} \\
\hline
\hline
Beam &  & \multicolumn{3}{c|}{Target} & \multicolumn{3}{c|}{Recoil} & \multicolumn{4}{c}{Target + Recoil} \\
  & -& -& -& -& $ x' $ & $ y' $ & $ z' $ & $ x' $ & $ x' $ & $ z' $ & $ z' $  \\
  & -& $ x $ & $ y $ & $ z $ & -& -& -& $ x $ & $ z $ & $ x $ & $ z $ \\
\hline
  &  &  &  &  &  &  &  &  &   \\
unpolarized & $ \sigma_{0} $ &  & $ T $ &  &  & $ P $ &  & $T_{x'}$ & $L_{x'}$ & $T_{z'}$ & $L_{z'}$ \\
  &  &  &  &  &  &  &  &  &   \\
linearly pol. & $ \Sigma $ & $ H $ & $ P $ & $ G $ & $ O_{x'} $ & $ T $ & $ O_{z'} $ &  &  &  &  \\
  &  &  &  &  &  &  &  &  &   \\
circularly pol. &  & $ F $ &  & $ E $ & $ C_{x'} $ &  & $ C_{z'} $  &  &  &  &  \\
\hline
\hline
\end{tabular} \label{tab:SandorfiObservableTable}}
\end{table}
in the CMS. The 4 functions $\left\{F_{i} \left(W,\theta\right); i = 1,\ldots,4\right\}$ are called CGLN-amplitudes and contain all information on the dynamics of the reaction. \newline
Since all particles in the reaction except for the meson $\mathcal{P}$ have spin, the preparation of the spin degrees of freedom in the initial state as well as the (generally more difficult) measurement of the polarization of the recoil baryon $B$ facilitate the experimental determination of 16 polarization observables, summarized in Table \ref{tab:SandorfiObservableTable}. All observables are definable as asymmetries among different polarization states (see \cite{Sandorfi}). They contain the unpolarized differential cross section $\sigma_{0}$, the three single spin observables $\left\{\Sigma, T, P\right\}$ (corresponding to beam, traget and recoil polarization), as well as twelve double polarization observables which are divisible into the distinct classes of beam-target (BT), beam-recoil (BR) and target-recoil (TR) observables. \newline
Once the equations connecting the measurable observables to the model independent production amplitudes are worked out (reference \cite{Sandorfi} contains instructions on how to do this), it becomes apparent that all of these relations can be summarized by the relation
\begin{equation}
\check{\Omega}^{\alpha} = \frac{q}{k} \frac{1}{2} \sum \limits_{i,j=1}^{4} F_{i}^{\ast} \hat{A}^{\alpha}_{ij} F_{j} = \frac{q}{k} \frac{1}{2} \left< F \right| \hat{A}^{\alpha} \left| F \right> \mathrm{,} \quad \alpha = 1,\ldots,16 \mathrm{.} \label{eq:BCGLNPForm}
\end{equation}
The 16 real profile functions $\check{\Omega}^{\alpha}$, connected to the polarization observables via $\check{\Omega}^{\alpha} = \sigma_{0} \Omega^{\alpha}$, are bilinear hermitean forms in the CGLN amplitudes and can be represented by the generally complex hermitean matrices $\hat{A}^{\alpha}$ (cf. \cite{MyDiplomaThesis} for a listing of those). \newline
A change of the basis of spin quantization for the photoproduction reaction allows for the definition of different systems of spin amplitudes. \newpage Helicity amplitudes $H_{i}\left(W,\theta\right)$ or transversity amplitudes $b_{i}\left(W,\theta\right)$ are possible choices (cf. \cite{ChTab}). The different kinds of amplitudes are all related among each other  in a linear and invertible way. Therefore, they can be seen as fully  equivalent regarding their information content. The expressions for the polarization observables in the afore mentioned different systems of spin amplitudes retain the mathematical structure of equation (\ref{eq:BCGLNPForm}), while the observables are now represented by different matrices
\begin{equation}
\check{\Omega}^{\alpha} = \frac{q}{k} \frac{1}{2} \left< H \right| \Gamma^{\alpha} \left| H \right> = \frac{q}{k} \frac{1}{2} \left< b \right| \tilde{\Gamma}^{\alpha} \left| b \right> \mathrm{.} \label{eq:BHTPForm}
\end{equation}
The $\Gamma^{\alpha}$ (or $\tilde{\Gamma}^{\alpha}$ in case of transversity amplitudes) are a set of 16 hermitean unitary Dirac $\Gamma$-matrices (cf. \cite{ChTab, MyDiplomaThesis}). They have useful properties, the exploitation of which facilitates the identification of complete experiments.

\section{Complete experiments for spin amplitudes} \label{sec:CompExChTab}

Since photoproduction allows access to 16 polarization observables but needs 4 complex amplitudes for a model independent description (constituting just 8 real numbers), the fact can be anticipated that measuring all observables would mean an overdetermination for the problem of extracting amplitudes. \newline
This issue has triggered investigations on so called complete experiments (cf. \cite{BarDo, ChTab}), which are subsets of a minimum number of observables that allow for a unique extraction of the amplitudes. Here one generally means unique only only up to an overall phase, since equations (\ref{eq:BCGLNPForm}, \ref{eq:BHTPForm}) are invariant by a simultaneous rotation of all amplitudes by the same phase. Also, the complete experiment problem is first of all a precise mathematical problem disgregarding measurement uncertainties. \newline
Chiang and Tabakin have published a solution to this problem (cf. \cite{ChTab}) that shall be depicted here. First of all it was noted that, using the fact that the $\tilde{\Gamma}$-matrices are an orthonormal basis of the complex $4 \times 4$-matrices, equation (\ref{eq:BHTPForm}) can be inverted in order to yield expressions for the bilinear products
\begin{equation}
b_{i}^{\ast} b_{j} = \frac{1}{2} \sum_{\alpha} \left(\tilde{\Gamma}^{\alpha}_{ij}\right)^{\ast} \check{\Omega}^{\alpha} \mathrm{.} \label{eq:BilinearBProducts}
\end{equation}
This relation allows for the determination of the moduli $\left|b_{i}\right|$ and relative phases $\phi^{b}_{ij}$ of the $b_{i}$ and therefore fully constrains them up to an overall phase. Generalizations of equation (\ref{eq:BilinearBProducts}) for helicity and CGLN amplitudes are possible (see \cite{MyDiplomaThesis}) but shall not be quoted here. \newpage
Another important property of the $\tilde{\Gamma}$ is that they imply quadratic relations among the observables known as the Fierz identities (see \cite{ChTab}) 
\begin{equation}
\check{\Omega}^{\alpha} \check{\Omega}^{\beta} = \sum_{\delta, \eta} C^{\alpha \beta}_{\delta \eta} \check{\Omega}^{\delta} \check{\Omega}^{\eta} \mathrm{,} \label{eq:FierzRelations}
\end{equation}
where $C^{\alpha \beta}_{\delta \eta} = \left( 1/16 \right) \mathrm{Tr} \left[\tilde{\Gamma}^{\delta} \tilde{\Gamma}^{\alpha} \tilde{\Gamma}^{\eta} \tilde{\Gamma}^{\beta}\right]$. \newline
Equations (\ref{eq:BilinearBProducts}) and (\ref{eq:FierzRelations}) are all that is needed to prove that 8 carefully chosen observables suffice in order to obtain a complete experiment (\cite{ChTab}). Among those should be the unpolarized cross section and the three single polarization observables.
The remaining quantities have to be picked from at least two different classes of double polarization observables, with no more than two of them from the same class. The word 'prove' means in this case that for all cases mentioned in reference \cite{ChTab}, equation (\ref{eq:FierzRelations})
was used to express the missing 8 observables in terms of the measured ones. \newline
In practical investigations of photoproduction data, the goal is not to determine the full reaction amplitudes, but rather the partial waves, in this case called multipoles.

\section{Complete experiments in a truncated partial wave analysis} \label{sec:CompExTPWA}

The expansions of the full amplitudes $F_{i}$ into multipoles are known (cf. eg. \cite{Sandorfi}). In case these expansions are truncated at some finite angular momentum quantum number $\ell_{\mathrm{max}}$, an approximation that is justified for reactions with supressed background contributions (eg. $\pi^{0}$ photoproduction), then the profile functions defined in equation (\ref{eq:BCGLNPForm}) can be arranged as a finite expansion into associated Legendre polynomials%
\begin{align}
\check{\Omega}^{\alpha} \left( W, \theta \right) &= \frac{q}{k} \hspace*{3pt} \sum \limits_{k = \beta_{\alpha}}^{2 \ell_{\mathrm{max}} + \beta_{\alpha} + \gamma_{\alpha}} \left(a_{L}\right)_{k}^{\alpha} \left( W \right) P^{\beta_{\alpha}}_{k} \left( \cos \theta \right) \mathrm{,}  \label{eq:LowEAssocLegParametrization1} \\
\left(a_{L}\right)_{k}^{\alpha} \left( W \right) &= \left< \mathcal{M}_{\ell_{\mathrm{max}}} \left( W \right) \right| \left( \mathcal{C}_{L}\right)_{k}^{\alpha} \left| \mathcal{M}_{\ell_{\mathrm{max}}} \left( W \right) \right> \mathrm{.} \label{eq:LowEAssocLegParametrization2}
\end{align}
The parameters $\beta_{\alpha}$ and $\gamma_{\alpha}$ defining this expansion are given in Table \ref{tab:AngularDistributions} (the whole notation is according to \cite{LotharNStar}). \newline
The real Legendre coefficients $\left(a_{L}\right)_{k}^{\alpha}$ are given as bilinear hermitean forms  in terms of the $4 \ell_{\mathrm{max}}$ multipoles, which are gathered in the vector $\left| \mathcal{M}_{\ell_{\mathrm{max}}} \right>$. Therefore, the problem of multipole-extraction from a set of fitted coefficients \newpage
\begin{table}[ht]
\centering
\caption{The parameters defining the TPWA problem, equations (\ref{eq:LowEAssocLegParametrization1}) and (\ref{eq:LowEAssocLegParametrization2}).}
{\begin{tabular}{ccc|ccccc||ccc|ccccccc}
\hline
\hline
  &  &  &  &  &  &  &  &  &  &  &  &  &  &   \\
Type & $ \check{\Omega}^{\alpha} $ & $\alpha$ &  & $ \beta_{\alpha} $ &  & $ \gamma_{\alpha} $ &  & Type & $ \check{\Omega}^{\alpha} $ & $\alpha$ &  & $ \beta_{\alpha} $ &  & $ \gamma_{\alpha} $ &   \\
\hline
  &  &  &  &  &  &  &  &  &  &  &  &  &  &   \\
  & $ I \left( \theta \right) $ & $1$ &  & $ 0 $ &  & $ 0 $ &  &  & $ \check{O}_{x'} $ & $14$ &  & $ 1 $ &  & $ 0 $ &   \\
 S & $ \check{\Sigma} $ & $4$ &  & $ 2 $ &  & $ -2 $ &  & BR & $ \check{O}_{z'} $ & $7$ &  & $ 2 $ &  & $ -1 $ &    \\
  & $ \check{T} $ & $10$ &  & $ 1 $ &  & $ -1 $ &  &  & $ \check{C}_{x'} $ & $16$ &  & $ 1 $ &  & $ 0 $ &    \\
  & $ \check{P} $ & $12$ &  & $ 1 $ &  & $ -1 $ &  &  & $ \check{C}_{z'} $ & $2$ &  & $ 0 $ &  & $ +1 $ &    \\
\hline
  &  &  &  &  &  &  &  &  &  &  &  &  &  &    \\
  & $ \check{G} $ & $3$ &  & $ 2 $ &  & $ -2 $ &  &  & $ \check{T}_{x'} $ & $6$ &  & $ 2 $ &  & $ -1 $ &    \\
 BT & $ \check{H} $ & $5$ &  & $ 1 $ &  & $ -1 $ &  & TR & $ \check{T}_{z'} $ & $13$ &  & $ 1 $ &  & $ 0 $ &   \\
  & $ \check{E} $ & $9$ &  & $ 0 $ &  & $ 0 $ &  &  & $ \check{L}_{x'} $ & $8$ &  & $ 1 $ &  & $ 0 $ &   \\
  & $ \check{F} $ & $11$ &  & $ 1 $ &  & $ -1 $ &  &  & $ \check{L}_{z'} $ & $15$ &  & $ 0 $ &  & $ +1 $ &   \\
\hline
\hline
\end{tabular} \label{tab:AngularDistributions}}
\end{table}
%
$\left(a_{L}\right)_{k}^{\alpha}$ leads to a similar mathematical structure compared to the equation (\ref{eq:BCGLNPForm}) encountered in the investigation of complete experiments in section \ref{sec:CompExChTab}. The question for such complete sets can now be asked again, but in the context of a truncated partial wave analysis (TPWA). \newline
It is a very interesting fact that in this case, the number of observables that is needed for completeness reduces as compared to the case with full production amplitudes. This is true at least in the mathematically precise situation, without measurement uncertainty. The algebra that is needed to prove this result was first worked out by Omelaenko \cite{Omelaenko} (for a recent and more detailed account, cf. \cite{OmelaenkoRevisited}). \newline
It is sufficient to investigate the discrete ambiguities allowed by the group S observables, i.e. $\left\{\sigma_{0}, \Sigma, T, P\right\}$.
It is then seen that the latter are invariant under one mathematical ambiguity transformation, called the 'double ambiguity', which is present in principle for all energies. There may also be additional pairs of solutions, called accidential ambiguities, depending on the numerical confguration of the Legendre coefficients. It can however be shown that those play no role for the mathematically exact case. The above mentioned double ambiguity on the other hand can be resolved by either the $F$ or $G$ observable, as well as every observable from the BR and TR classes. Therefore one is lead to mathematically complete sets containing just 5 observables, for example
\begin{equation}
\left\{ \sigma_{0}, \Sigma, T, P, F \right\} \mathrm{.} \label{eq:CompSetTPWAExample}
\end{equation}

\newpage

\section{TPWA fits using the bootstrapping method} \label{sec:MATHFits}

Here we will describe preliminary results of a TPWA fit to actual data comprising the set of observables (\ref{eq:CompSetTPWAExample}).
The observables $\sigma_{0}$ and $\Sigma$ are taken from the works \cite{MAMIData} and \cite{LeukelPhD}. Recent measurements of $T$
and $F$ were performed at MAMI \cite{SchumannOtteData}. For $P$ we take the Kharkov data \cite{BelyaevData}. \newline
The fit procedure proceeds as follows, using a truncation at $\ell_{\mathrm{max}}=1$ ($S$- and $P$-waves). First, Legendre coefficients are determined by fitting the angular distributions (\ref{eq:LowEAssocLegParametrization1}) to the data. The index set for the fitted observables is $\alpha_{F} \in \left\{1,4,10,12,11\right\}$ (cf. Table \ref{tab:AngularDistributions}) in this particular case here. In the next step, we minimize the functional (up to now omitting correlations)
\begin{equation}
\Phi_{\mathcal{M}}\left(\mathcal{M}_{\ell_{\mathrm{max}}}\right) = \sum_{\alpha_{F}, k} \left( \frac{\left(a_{L}^{\mathrm{Fit}}\right)_{k}^{\alpha_{F}} - \left< \mathcal{M}_{\ell_{\mathrm{max}}} \right| \left( \mathcal{C}_{L}\right)_{k}^{\alpha_{F}} \left| \mathcal{M}_{\ell_{\mathrm{max}}} \right>}{\Delta \left(a_{L}^{\mathrm{Fit}}\right)_{k}^{\alpha_{F}}} \right)^{2} \mathrm{,} \label{eq:Chi2Functional}
\end{equation}
using the results from the angular fit and varying the real and imaginary parts of the multipoles (the FindMinimum routine of MATHEMATICA is employed). The overall phase of the multipoles is constrained to $\mathrm{Re}\left[E_{0+}\right] \geq 0$ and $\mathrm{Im}\left[E_{0+}\right] = 0$, since this phase can never be obtained from a truncated fit to the data alone. \newline
In order to exclude any kind of model dependencies, the start parameters for the fit are not taken from a prediction, but are determined randomly by using a Monte Carlo sampling of the relevant, $\left(8 \ell_{\mathrm{max}} -1\right) = 7$ dimensional multipole space (the space spanned by the real and imaginary parts). This sampling is simplified by the fact that the total cross section $\hat{\sigma}$, being a sum of moduli-squared of multipoles, already constrains the relevant part of the multipole space to a 6 dimensional ellipsoid.
\begin{figure}[h]
\centering
 \begin{overpic}[width=0.2425\textwidth]%
      {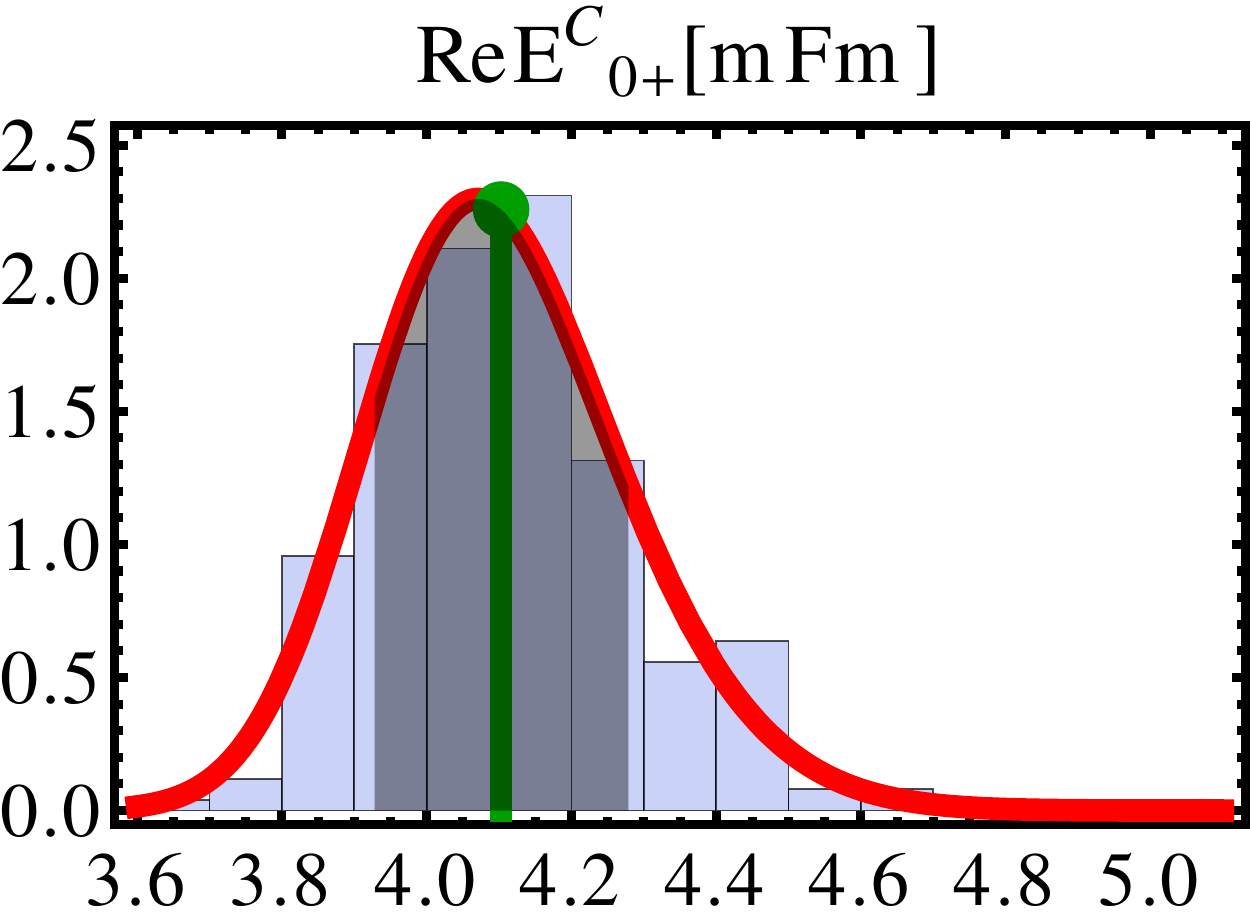}
    \put(26.5,18.5){\rotatebox{30}{\textcolor{gray}{Preliminary}}}
\end{overpic}
\begin{overpic}[width=0.2425\textwidth]%
      {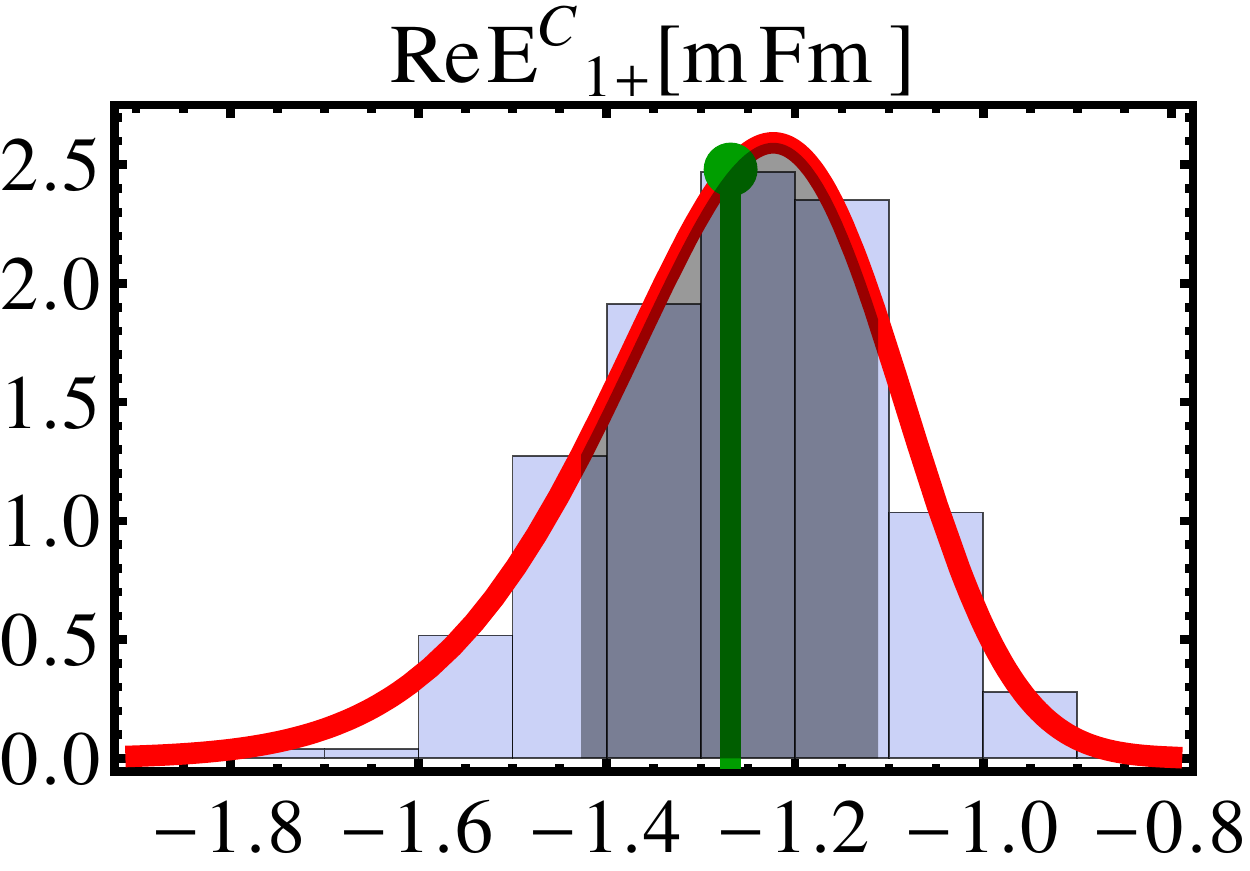}
    \put(26.5,18.5){\rotatebox{30}{\textcolor{gray}{Preliminary}}}
\end{overpic}
\begin{overpic}[width=0.2425\textwidth]%
      {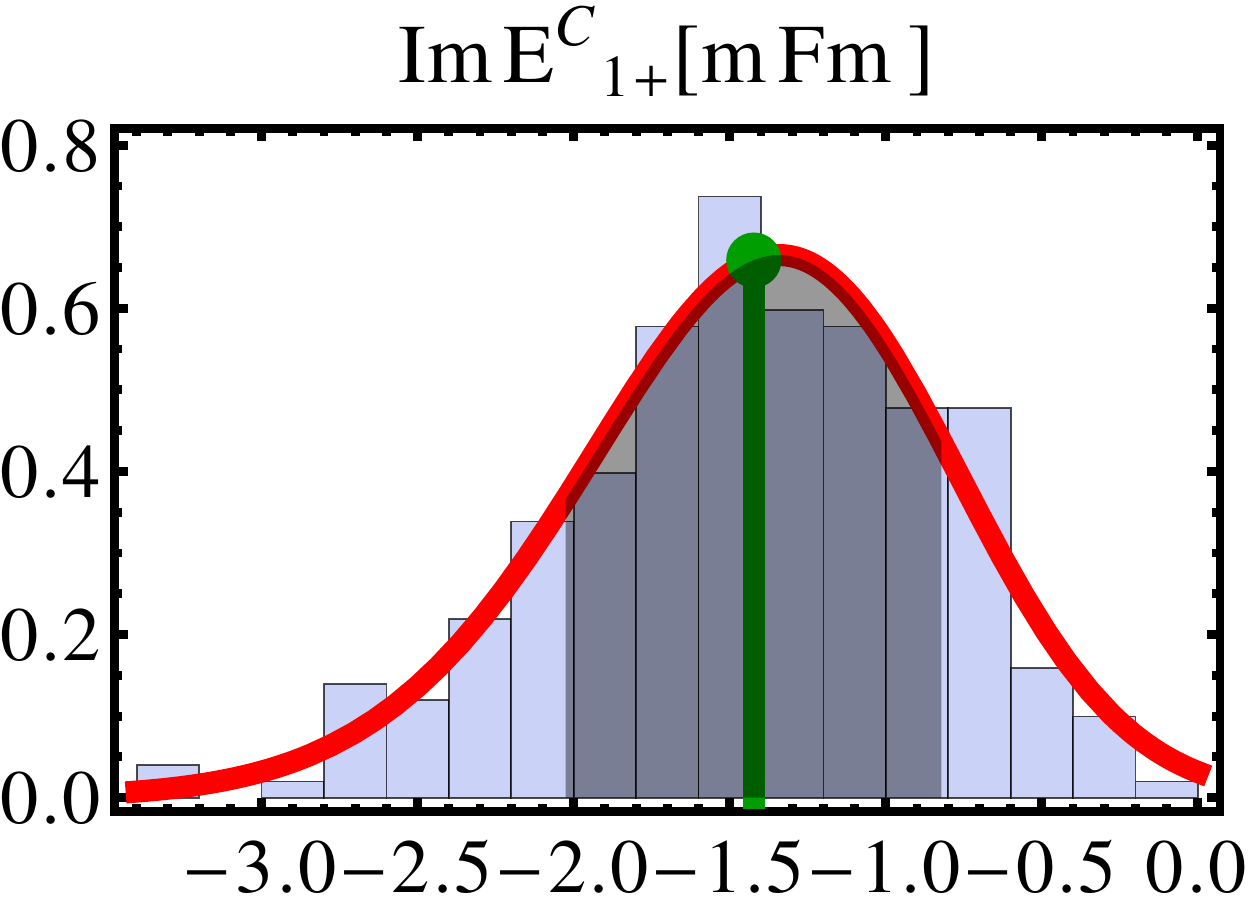}
    \put(26.5,18.5){\rotatebox{30}{\textcolor{gray}{Preliminary}}}
\end{overpic}
\begin{overpic}[width=0.2425\textwidth]%
      {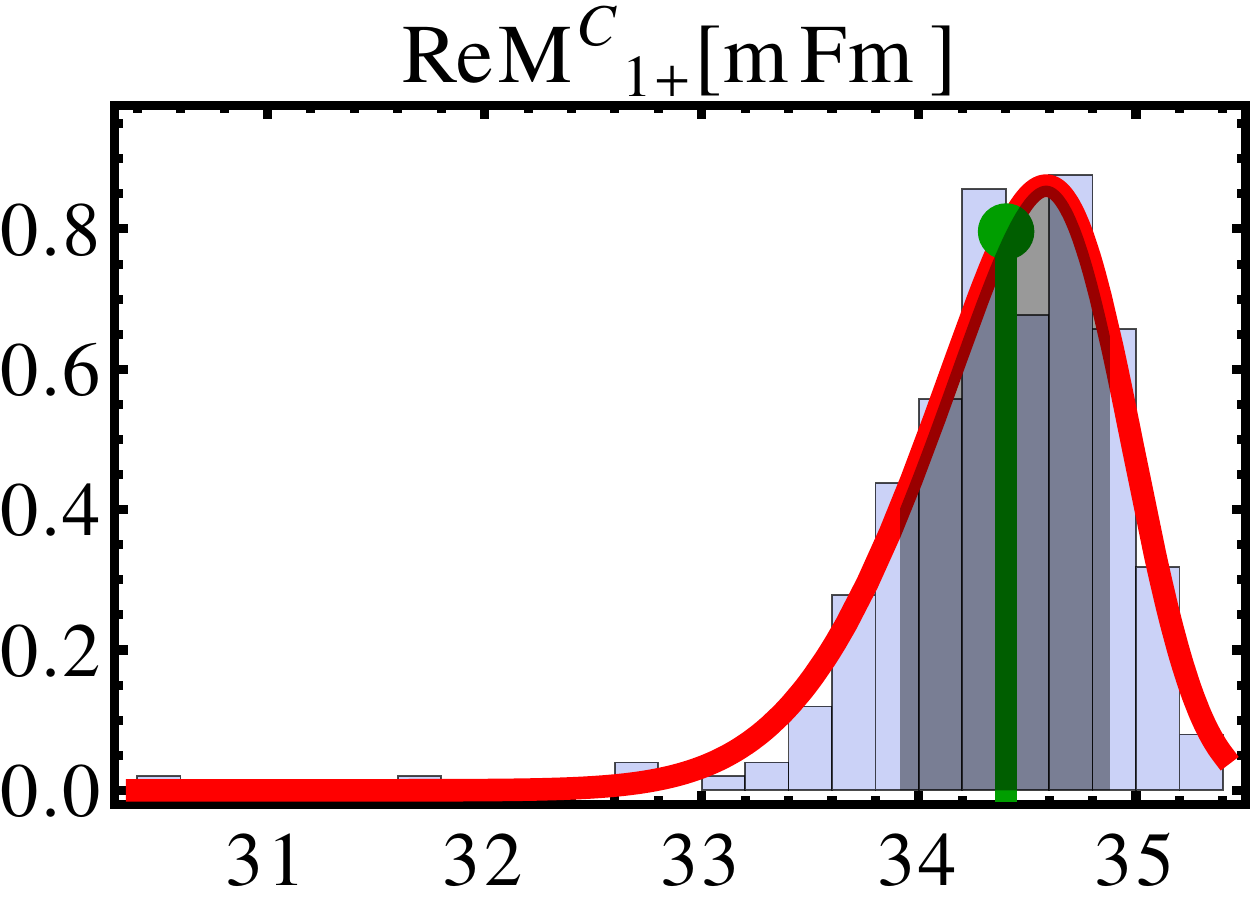}
    \put(26.5,18.5){\rotatebox{30}{\textcolor{gray}{Preliminary}}}
\end{overpic} \\
\vspace*{2pt}
\begin{overpic}[width=0.2425\textwidth]%
      {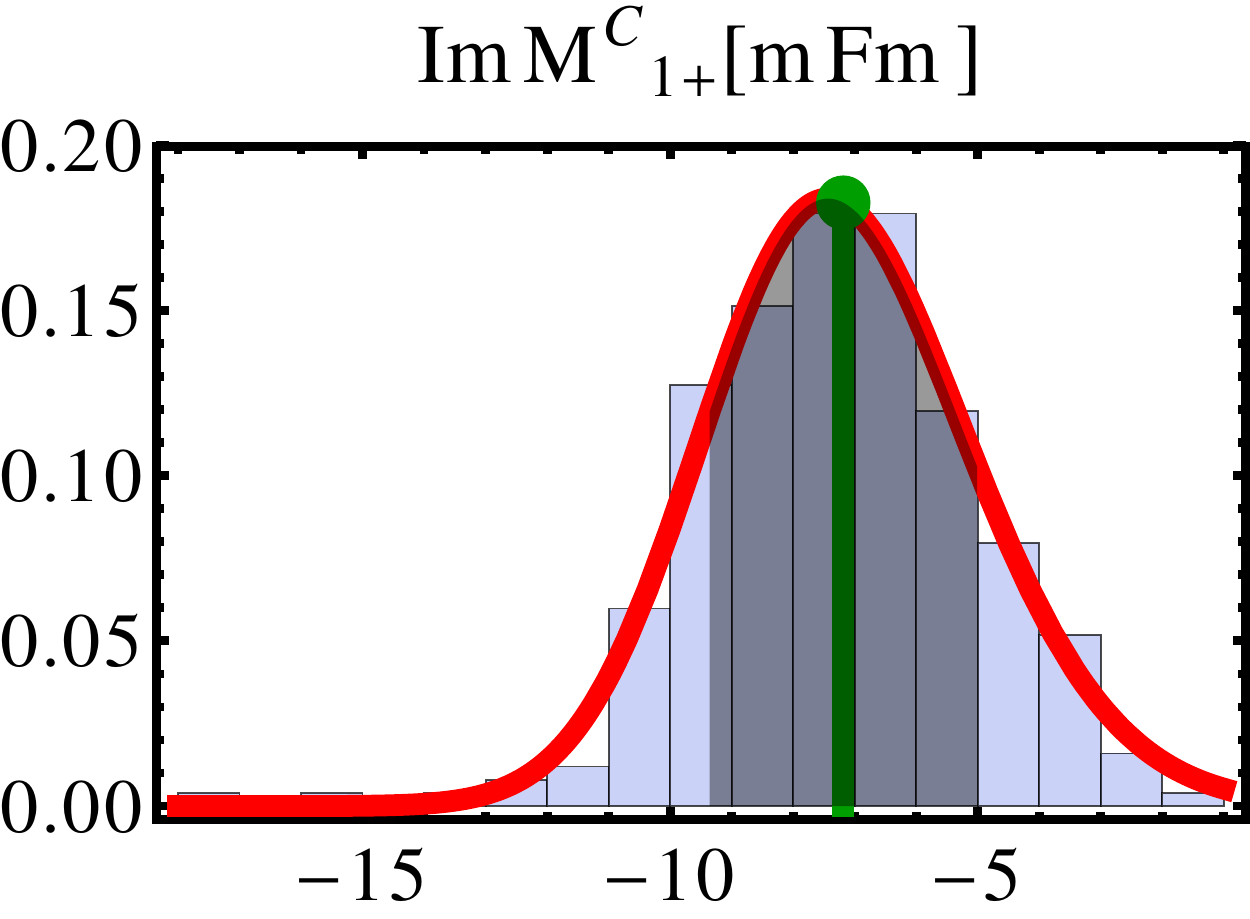}
    \put(26.5,18.5){\rotatebox{30}{\textcolor{gray}{Preliminary}}}
\end{overpic}
\begin{overpic}[width=0.2425\textwidth]%
      {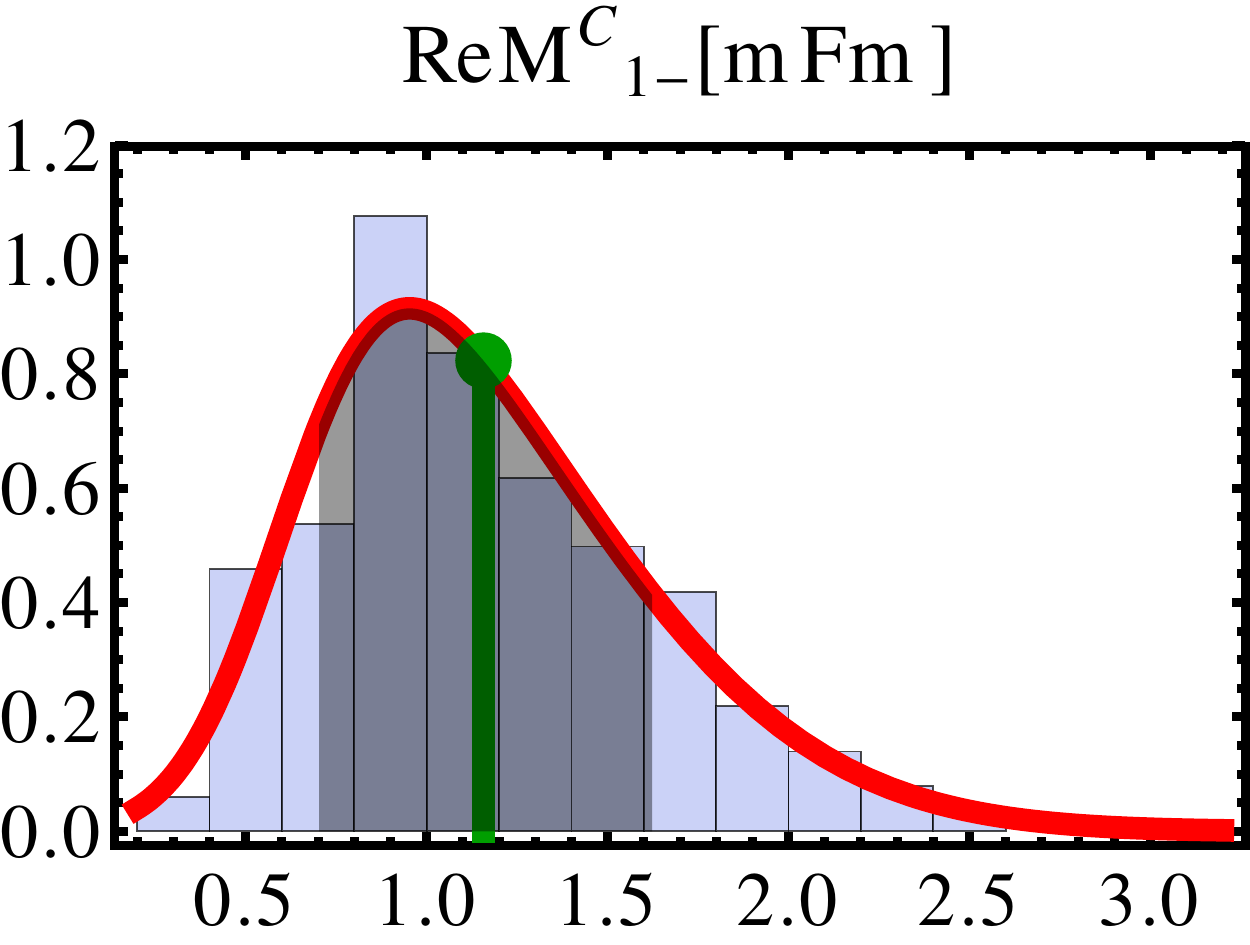}
    \put(26.5,18.5){\rotatebox{30}{\textcolor{gray}{Preliminary}}}
\end{overpic}
\begin{overpic}[width=0.2425\textwidth]%
      {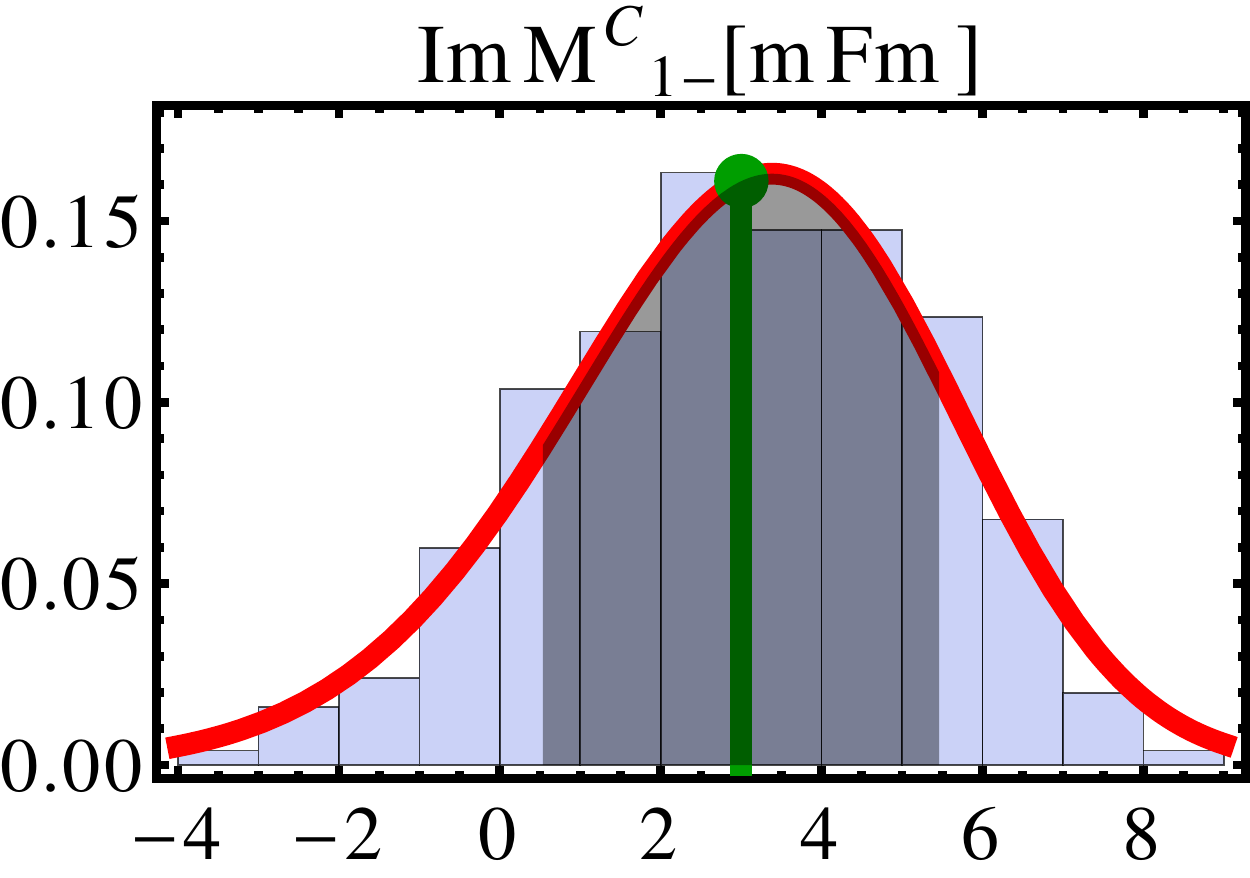}
    \put(26.5,18.5){\rotatebox{30}{\textcolor{gray}{Preliminary}}}
\end{overpic}
\caption{Histograms resulting from the bootstrapping at $E_{\gamma}^{\mathrm{LAB}} \approx 338 \hspace*{1pt} \mathrm{MeV}$.}
\label{fig:Histograms}
\end{figure}
\newpage
%
\vspace*{-10pt}
\begin{figure}[h]
\centering
 \begin{overpic}[width=0.3275\textwidth]%
      {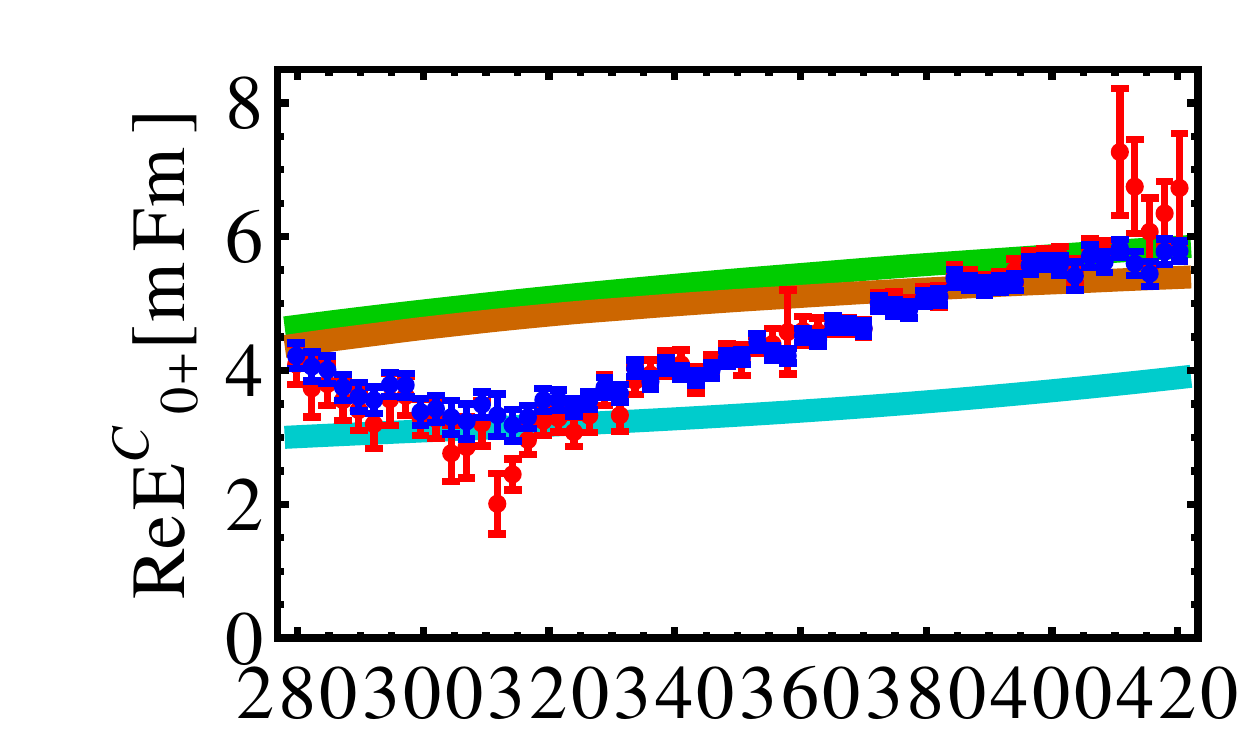}
    \put(36,16){\rotatebox{30}{\textcolor{gray}{Preliminary}}}
\end{overpic}
\begin{overpic}[width=0.3275\textwidth]%
      {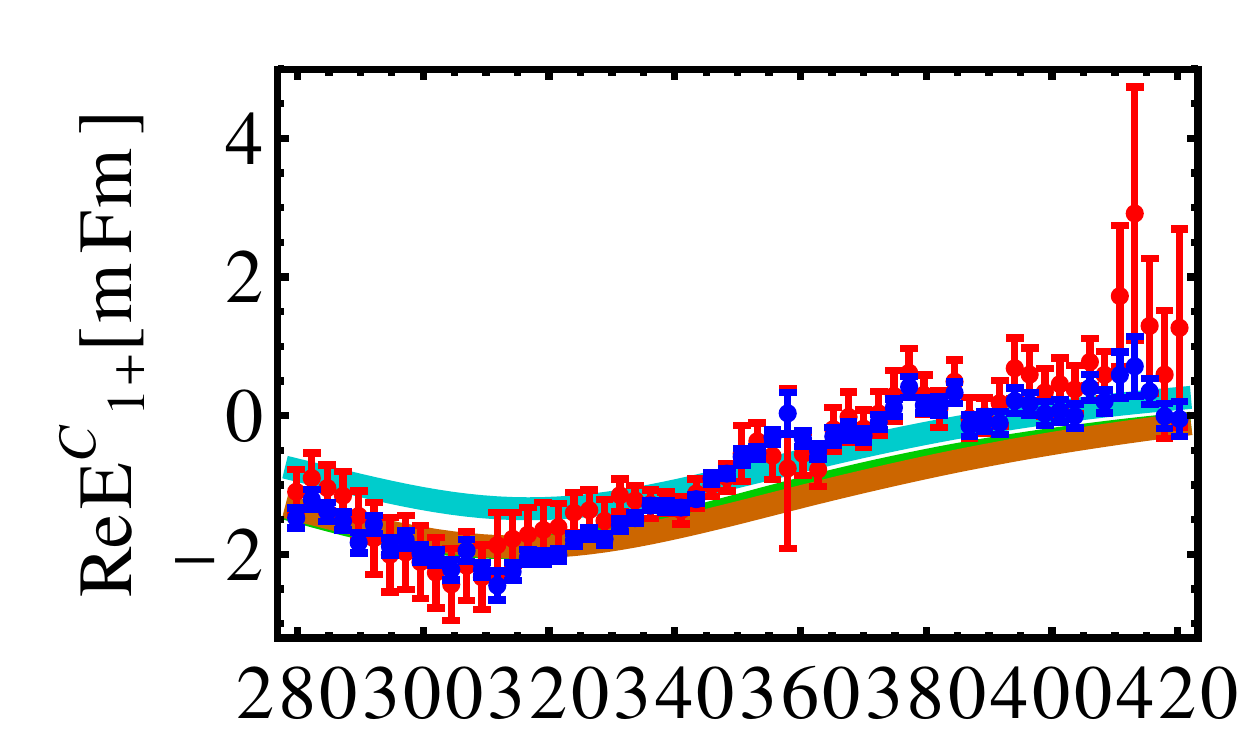}
    \put(36,16){\rotatebox{30}{\textcolor{gray}{Preliminary}}}
\end{overpic}
\begin{overpic}[width=0.3275\textwidth]%
      {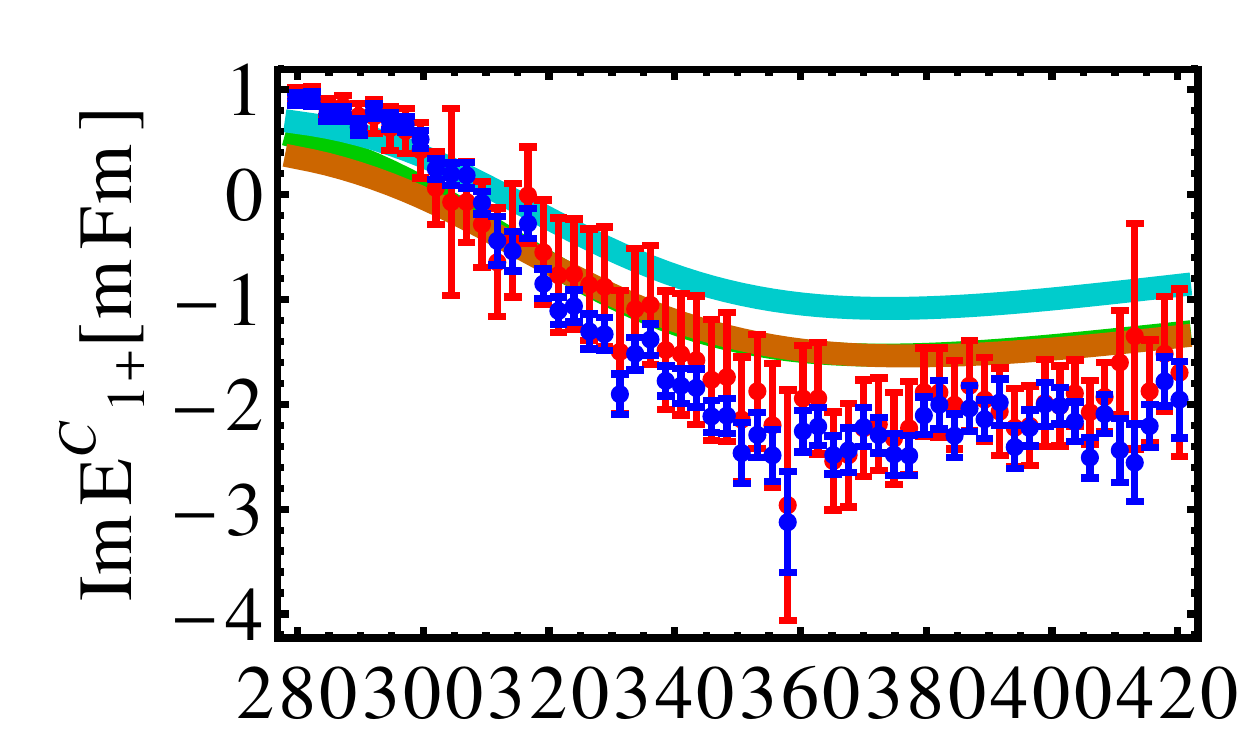}
    \put(36,16){\rotatebox{30}{\textcolor{gray}{Preliminary}}}
\end{overpic} \\
\begin{overpic}[width=0.3275\textwidth]%
      {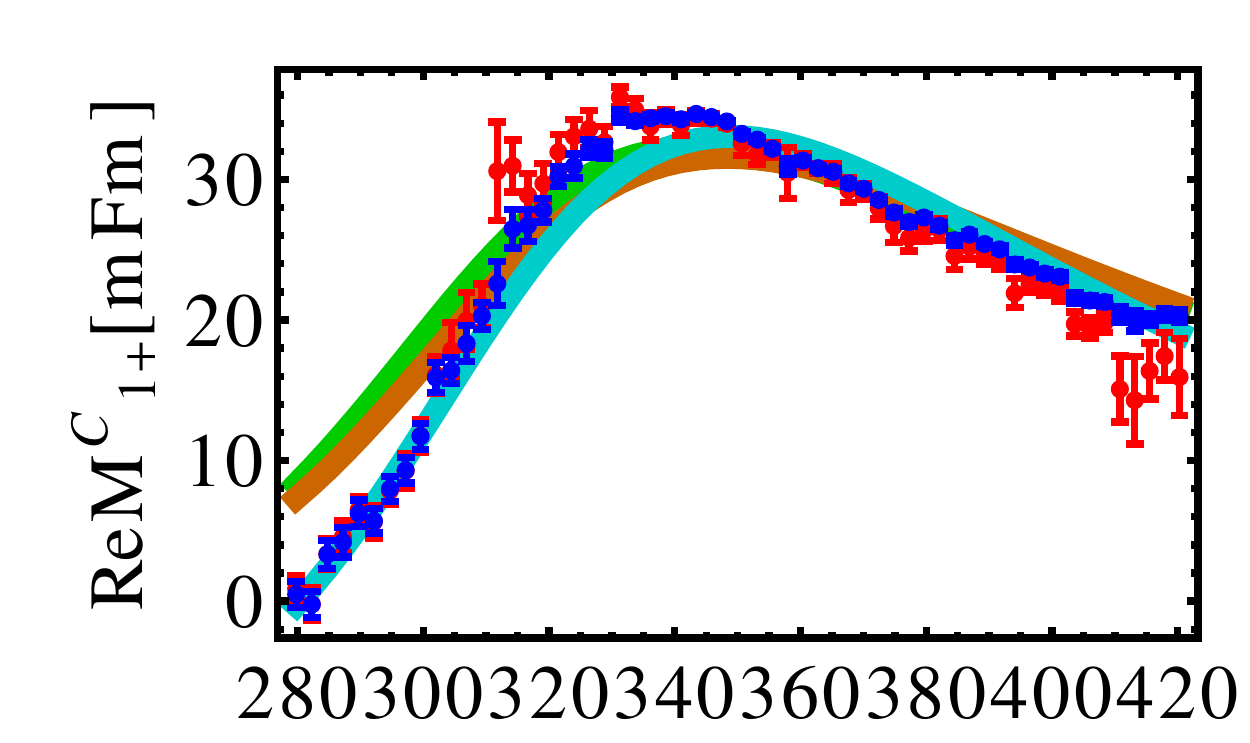}
    \put(36,16){\rotatebox{30}{\textcolor{gray}{Preliminary}}}
    \put(35,-10){$E_{\gamma}^{\mathrm{LAB}}$\hspace*{1.5pt}[MeV]}
\end{overpic}
\begin{overpic}[width=0.3275\textwidth]%
      {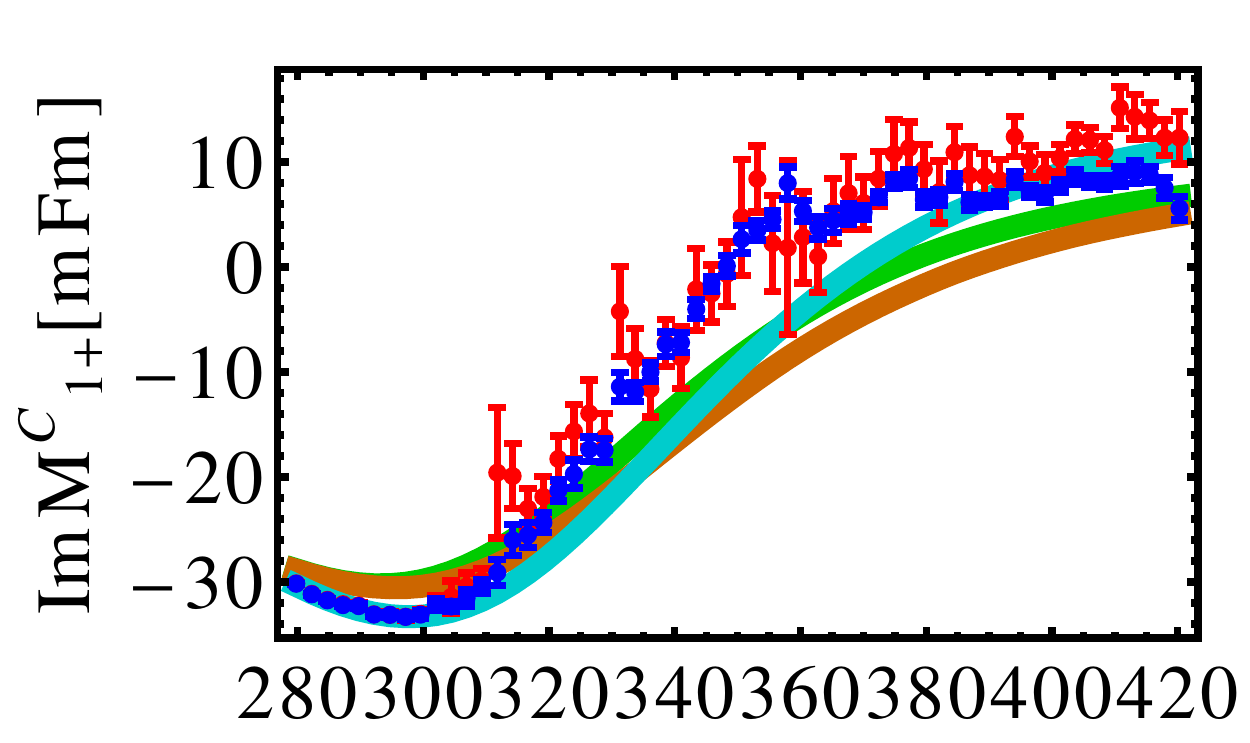}
    \put(36,16){\rotatebox{30}{\textcolor{gray}{Preliminary}}}
\end{overpic}
\begin{overpic}[width=0.3275\textwidth]%
      {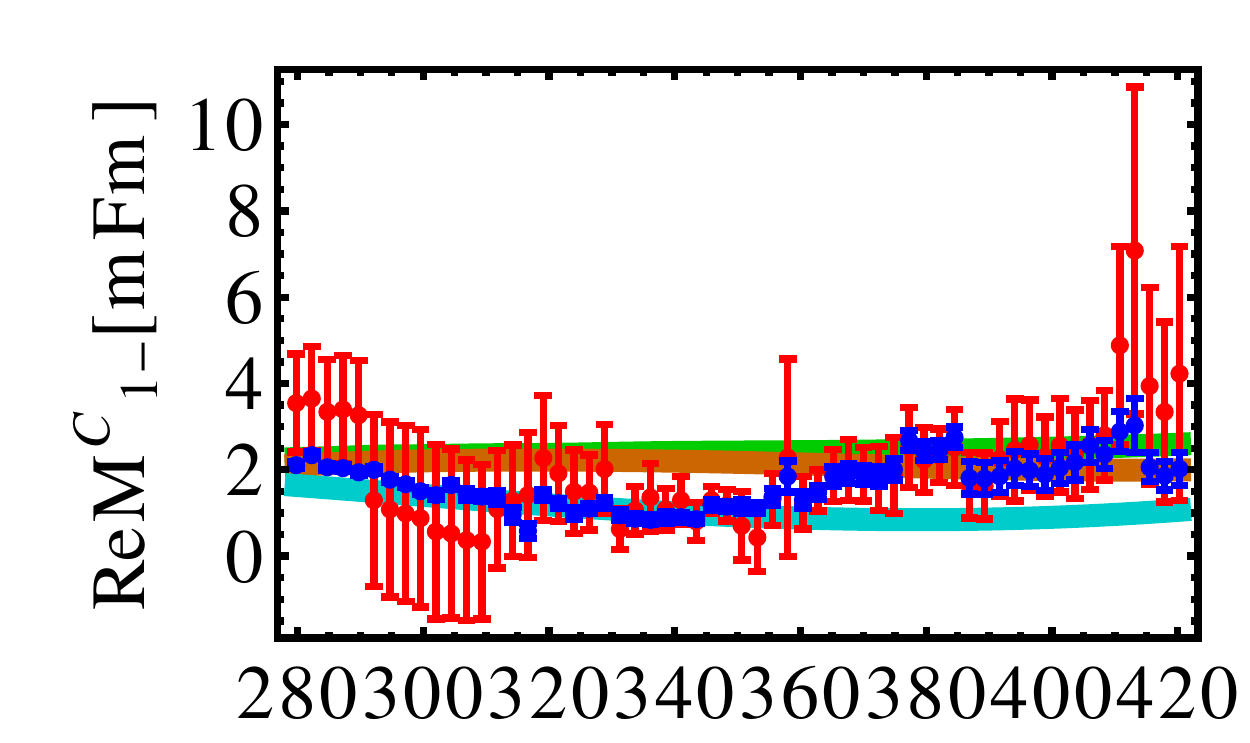}
    \put(36,16){\rotatebox{30}{\textcolor{gray}{Preliminary}}}
    \put(35,-10){$E_{\gamma}^{\mathrm{LAB}}$\hspace*{1.5pt}[MeV]}
\end{overpic} \\
\begin{overpic}[width=0.3275\textwidth]%
      {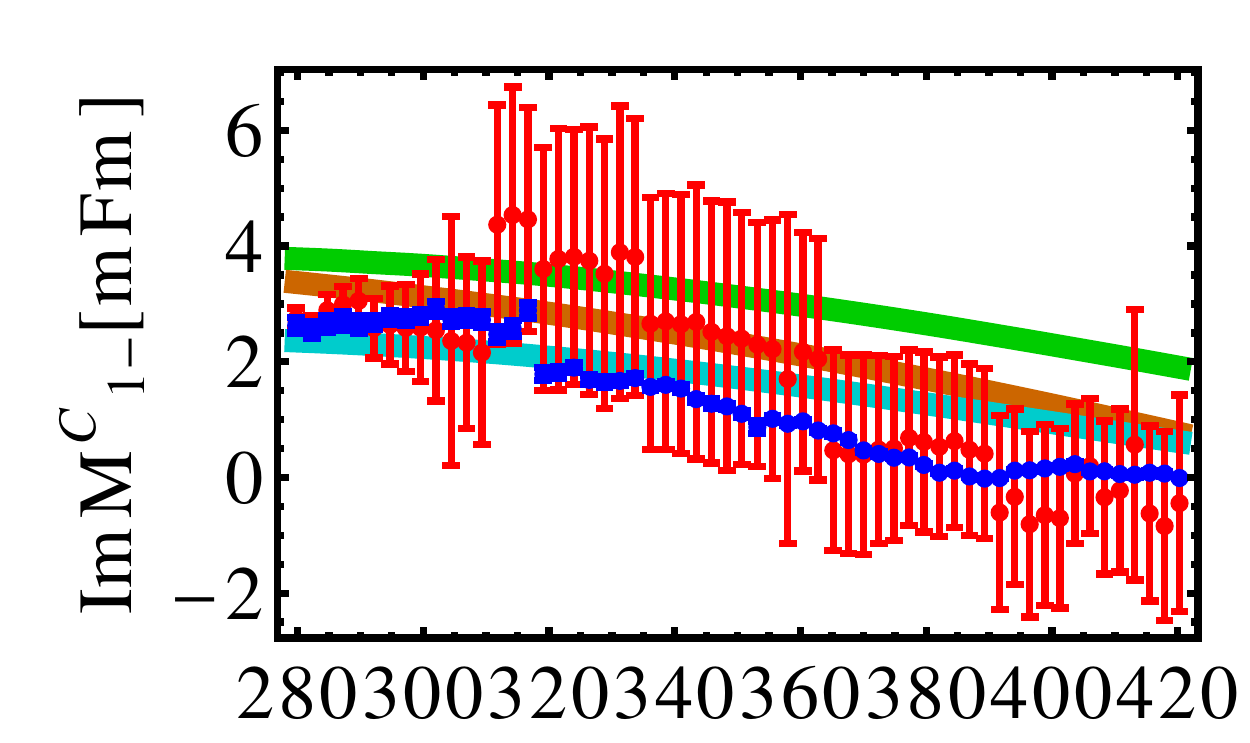}
    \put(36,16){\rotatebox{30}{\textcolor{gray}{Preliminary}}}
\end{overpic}
\caption{Results of the bootstrapping procedure for $S$- and $P$-wave multipoles (red: Kharkov $P$ data; blue: SAID-predictions for $P$). The colored model curves are for comparison taken from MAID \cite{MAID} (green), SAID \cite{SAID} (brown) and Bonn-Gatchina \cite{BoGa} (cyan).}
\label{fig:MultipoleResults}
\end{figure}
The amount of $N_{MC} = 1000$ Monte Carlo start configurations was chosen. It has to be reported that using this procedure, it was possible to find a pronounced best minimum for the dataset under investigation. \newline
In addition one would wish to have an estimate for the errors of the resulting multipoles, as well as a check whether the data allow any ambiguities caused by their finite precision. To achieve both tasks, a method known as 'bootstrapping' was chosen (\cite{BootstrapEfronPaper1977}). In this approach, the data are resampled using a gaussian distribution function centered at $\mu = \check{\Omega}^{\alpha}$ having a width $\sigma = \Delta \check{\Omega}^{\alpha}$ for each datapoint. In this way, an ensemble of $250$ additional datasets was generated, each time starting at the original datapoints. The above mentioned TPWA fit procedure was then applied to each ensemble member. If a good minimum is found in each case, one can histogram the results and extract mean and width for each parameter (cf. Figure \ref{fig:Histograms}). \newline
The bootstrap did not show any indications of ambiguities allowed by the data. Therefore, the results for mean and width of the single solution found can be plotted against energy, the result of which is shown in Figure \ref{fig:MultipoleResults}. 
\newpage
Because the errors of the Kharkov dataset are very large, additional fits were performed replacing these data by a SAID-prediction for $P$ which has been endowed with a $5\%$-error at each datapoint. The results indicate that the uncertainty of the multipoles, especially for $M_{1-}$, is quite sensitive to this replacement (Figure \ref{fig:MultipoleResults}).

\section{Summary and outlook} \label{sec:Summary}

Mathematically complete sets of observables contain a minimum number of 8 in the case of spin amplitude extraction and 5 for a TPWA. First investigations of a particularly simple fit in the $\Delta$-region confirm the latter result. \newline
Bootstrapping methods were proposed in order to get a good estimate for the error of the fitted multipoles. This error is seen to shrink in case more precise pseudodata for the recoil polarization observable $P$ are introduced. \newline \newline

\hspace*{-10pt} {\bf Acknowledgments} \newline \newline
   The author wishes to thank the organizers for the hospitality, as well as for providing 
   a very relaxed and friendly atmosphere during the workshop. \newline
   This work was supported by the \textit{Deutsche Forschungsgemeinschaft} within SFB/TR16.

\newpage

\selectlanguage{english}


\end{document}